\documentclass[journal]{IEEEtran}
\usepackage{cite}
\usepackage{amsmath,amssymb,amsfonts}
\usepackage{algorithmic}
\usepackage{graphicx}
\usepackage{caption}
\usepackage{textcomp}
\usepackage{times}
\usepackage{bm}
\usepackage{multirow}
\usepackage{enumerate}
\usepackage{graphicx}
\usepackage{subfigure} 
\usepackage{multirow}
\usepackage{algorithmic}
\usepackage{algorithm}
\usepackage{color}
\usepackage{array}
\usepackage{epstopdf}
\definecolor{tabcolor}{rgb}{1,0,0}
\DeclareCaptionLabelSeparator{twospace}{.\ ~}
\def\BibTeX{{\rm B\kern-.05em{\sc i\kern-.025em b}\kern-.08em
		T\kern-.1667em\lower.7ex\hbox{E}\kern-.125emX}}
\begin{document} 
\title{Fluid Antenna-enabled Integrated Sensing, Communication, and Computing Systems}
\author{Yiping Zuo, Jiajia Guo, Weicong Chen, Weibei Fan, \\ Biyun Sheng, Fu Xiao, \IEEEmembership{\normalsize {Member,~IEEE}}, and Shi Jin, \IEEEmembership{\normalsize {Fellow,~IEEE}} 
\thanks{Yiping Zuo, Weibei Fan, Biyun Sheng, and Fu Xiao are with the School of Computer Science, Nanjing University of Posts and Telecommunications, Nanjing 210023, China (Email: zuoyiping@njupt.edu.cn, wbfan@njupt.edu.cn, biyunsheng@njupt.edu.cn, xiaof@njupt.edu.cn).}
\thanks{Jiajia Guo, Weicong Chen, and Shi Jin are with the National Mobile Communications Research Laboratory, Southeast University, Nanjing, 210096, China (Email: jiajiaguo@seu.edu.cn, cwc@seu.edu.cn jinshi@seu.edu.cn).}
}
\maketitle
\begin{abstract}
The current integrated sensing, communication, and computing (ISCC) systems face significant challenges in both efficiency and resource utilization. To tackle these issues, we propose a novel fluid antenna (FA)-enabled ISCC system, specifically designed for vehicular networks. We develop detailed models for the communication and sensing processes to support this architecture. An integrated latency optimization problem is formulated to jointly optimize computing resources, receive combining matrices, and antenna positions. To tackle this complex problem, we decompose it into three sub-problems and analyze each separately. A mixed optimization algorithm is then designed to address the overall problem comprehensively. Numerical results demonstrate the rapid convergence of the proposed algorithm. Compared with baseline schemes, the FA-enabled vehicle ISCC system significantly improves resource utilization and reduces latency for communication, sensing, and computation.
\end{abstract}


\IEEEpeerreviewmaketitle
  
\section{Introduction} \label{sec:Introduction}
Integrated sensing, communication, and computing (ISCC) systems are at the forefront of wireless networks \cite{li2023over}. Traditionally, sensing, communication, and computing tasks are handled by separate systems, leading to inefficiencies and increased latency. ISCC systems aim to integrate these functions into a unified framework, enabling more efficient resource utilization and faster response times\cite{ding2022joint,li2023integrated,liu2024joint}. Despite advancements in ISCC systems, current wireless communication systems still face significant challenges in both efficiency and resource utilization \cite{ma2019sensing}. Traditional systems, with their fixed antenna structures and separate handling of sensing, communication, and computing tasks, often result in suboptimal performance and increased operational costs\cite{ma2019sensing,gunlu2023secure,wei2022toward}. These studies highlight the need for innovative approaches to enhance the efficiency of ISCC systems. This prompts further research into integrating advanced technologies such as fluid antennas (FAs)\cite{wong2021fluid}, also known as movable antenna \cite{zhu2024movable1}.

FA technology is an emerging innovation in wireless communications, offering unparalleled flexibility and adaptability. Unlike traditional fixed antennas, fluid antennas can dynamically alter their physical and electrical properties. This allows them to adapt to changing environmental conditions and communication demands. This adaptability not only improves signal quality and reduces interference but also enhances overall network performance. FA technology is being explored in various scenarios, including fundamental communication tasks such as beamforming \cite{chen2024joint} and channel estimation\cite{psomas2023fluid}.  Additionally, FA is in combination with enabling technologies like MEC\cite{zuo2024fluid}, reconfigurable intelligent surfaces (RIS)\cite{ghadi2024performance}, and ISAC\cite{zou2024shifting}. Based on the existing research on FA, we hypothesize that incorporating FA technology into ISCC systems can significantly improve efficiency and resource utilization.

To the best of the authors’ knowledge, existing studies often overlook the potential synergies between ISCC and FA technologies, resulting in suboptimal solutions that fail to fully exploit their combined capabilities. In this paper, we first propose a novel FA-enabled ISCC system, applied in vehicular networks as a practical scenario. Specifically, we discuss detailed models for the communication and sensing processes in the proposed FA-enabled vehicle ISCC system. Then, we formulate an integrated latency optimization problem to jointly optimize computing resources, receive combining matrices, and antenna positions. To solve this complex problem, we decompose it into three sub-problems. We also develop a mixed optimization algorithm to find the optimal solutions. Through extensive simulations, we demonstrate the rapid convergence of the proposed  algorithm. The results show significant latency improvements in communication, sensing, and computation over baseline schemes. 

\section{System Model} \label{sec:SystemModel}
\begin{figure}[!t]
	\centering    
	\includegraphics[width=0.8\linewidth]{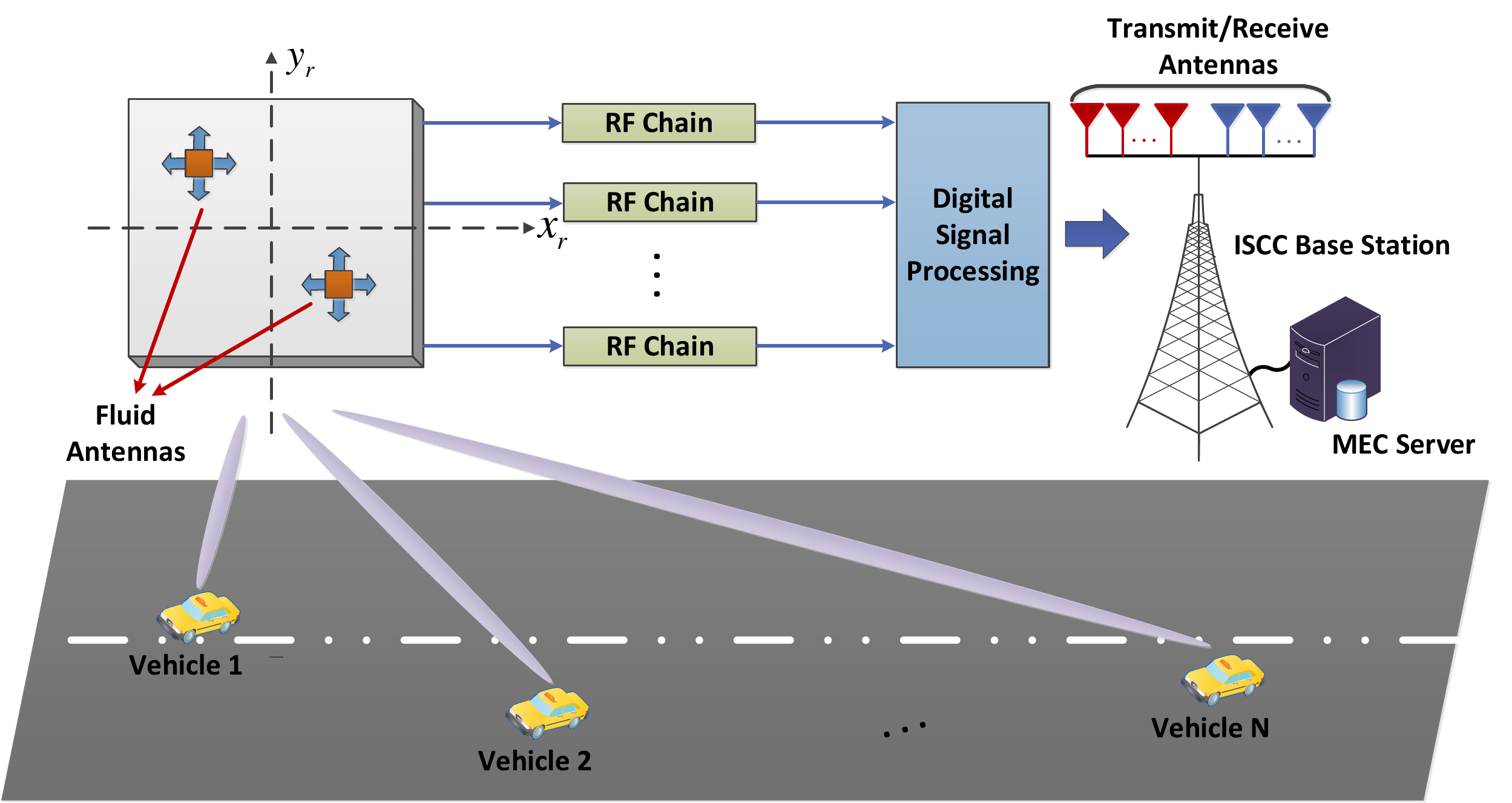} 
	\captionsetup{justification=justified}  
	\caption{The architecture of FA-enabled vehicle ISCC systems}
	\label{fig:SystemModel}
\end{figure}

As illustrated in Fig.~\ref{fig:SystemModel}, we propose an innovative FA-enabled vehicle ISCC system. This system features an ISCC BS and $N$ vehicles. The ISCC BS broadcasts a common signal to all vehicles and uses the echo signals for sensing their states, such as position and speed. In such a FA-enabled vehicle ISCC system, the transmitted signal is a dual-functional waveform for both radar sensing and communication. The ISCC BS is equipped with $M$ transmit/receive antennas to enhance communication and sensing performance, serving $N$ vehicles and simultaneously sensing their states. Each vehicle is equipped with a fixed-position antenna. Additionally, the ISCC BS incorporates a MEC server, which is managed by the cloud service provider of the core network. The set of all vehicles is denoted by ${\cal N} = \left\{ {1,2, \cdots ,N} \right\}$ and the set of all FAs at the ISCC BS is denoted by ${\cal M} = \left\{ {1,2, \cdots ,M} \right\}$. Notably, we assume that FAs on the ISCC BS are mobile with a localized domain, which is represented as a rectangle in a two-dimensional coordinate system, denoted as ${{\cal D}_r}$. Each FA is connected to the radio frequency (RF) chain through a flexible cable, which enhances the channel conditions between the ISCC BS and vehicles. We employ space-division multiple access to facilitate concurrent uplink communications between the vehicles and the ISCC BS. Consequently, we assume that the number of vehicles does not exceed the number of FAs at the ISCC BS, i.e., $N \le M$. The proposed system has $\mathcal{T}$ time periods, and we first focus on the performance in time period $t$, and then expand it to the entire timeline. Within each time slot, resource allocations such as antenna positioning, receive combining, and computing resources are assumed to remain constant. The position of the $m$-th receive FA at the ISCC BS is defined as ${{\mathbf{d}}_m}\left( t \right) = {\left[ {{x_m}\left( t \right),{y_m}\left( t \right)} \right]^{\text{T}}} \in {\mathcal{D}_r}$ for $m \in {\cal M}$ in time slot $t$.

\subsection{Communication Model}\label{subsec:CommunicationModel}
We consider the uplink transmission from vehicles to the ISCC BS. Then, the received signal ${{\bf{y}}_c}\left( t \right) \in {\mathbb{C}^{N \times 1}}$ at the ISCC BS in time slot $t$ is
\begin{equation}\label{ReceivedSignal_y} 
{{\bf{y}}_c}\left( t \right) = {\bf{W}}_c^H\left( t \right){{\bf{H}}_c}\left( {t,{\bf{d}}} \right){{\bf{P}}^{^{{1 \mathord{\left/ {\vphantom {1 2}} \right. \kern-\nulldelimiterspace} 2}}}}{\bf{s}}\left( t \right) + {\bf{W}}_c^H\left( t \right){{\bf{n}}_c}\left( t \right),
\end{equation}
where ${{\bf{W}}_c}\left( t \right) = \left[ {{{\bf{w}}_{1,c}}\left( t \right), \cdots ,{{\bf{w}}_{N,c}}\left( t \right)} \right] \in {\mathbb{C}^{M \times N}}$ represents the receiving combination matrix at the ISCC BS with ${{\bf{w}}_{n,c}}\left( t \right)$ being the combining vector for the transmitted signal of vehicle $n$. ${{\bf{H}}_c}\left( {t,{\bf{d}}} \right) = \left[ {{{\bf{h}}_{1,c}}\left( {t,{\bf{d}}} \right), \cdots ,{{\bf{h}}_{N,c}}\left( {t,{\bf{d}}} \right)} \right] \in {\mathbb{C}^{M \times N}}$ is the multiple-access channel matrix from all $N$ vehicles to the $M$ FAs at the ISCC BS with ${\bf{d}}\left( t \right) = {\left[ {{\bf{d}}_1^T\left( t \right), \cdots ,{\bf{d}}_M^T\left( t \right)} \right]^T}$ denoting the antenna position vector for FAs. ${{\bf{P}}^{{1 \mathord{\left/ {\vphantom {1 2}} \right. \kern-\nulldelimiterspace} 2}}} = {\rm{diag}}\left\{ {\sqrt {{p_1}} , \cdots ,\sqrt {{p_N}} } \right\} \in {\mathbb{C}^{N \times N}}$ denotes the power scaling matrix, where ${p_n}$ is the transmit power of vehicle $n$. ${\bf{s}}\left( t \right) = {[{s_1}\left( t \right), \cdots ,{s_N}\left( t \right)]^T} \in {\mathbb{C}^{N \times 1}}$ is the transmit signal vector of all vehicles, and ${s_n}\left( t \right)$ denotes the transmitted signal of vehicle $n$ with normalized power, i.e., $\mathbb{E}\left( {{\mathbf{s}}{{\mathbf{s}}^H}} \right) = {{\mathbf{I}}_N}$. Additionally, ${{\bf{n}}_c}\left( t \right) = {\left[ {{n_{1,c}}\left( t \right), \cdots ,{n_{M,c}}\left( t \right)} \right]^T} \sim \mathcal{C}\mathcal{N}\left( {{\mathbf{0}},{\sigma _c^2}{{\mathbf{I}}_M}} \right)$ denotes the zero-mean additive white Gaussian noise (AWGN) with ${\sigma _c^2}$ denoting the average noise power, in which $n_{m,c}$ is the noise at the $m$-th FA antenna at the ISCC BS.

The channel vector between each vehicle and the ISCC BS is shaped by the propagation environment and the location of FAs. Given that the movement range of FAs at the ISCC BS is significantly smaller compared to the overall signal propagation distance, it's reasonable to assume that the far-field condition holds between the vehicles and the ISCC BS. Consequently, for each vehicle, the angles of arrival (AoAs) and the magnitudes of the complex path coefficients across multiple channel paths remain invariant with respect to different FA positions, implying that only phase variations occur within the multiple channels in the reception area. Let $L_n$ denote the total number of receive channel paths at the ISCC BS from vehicle $n$. The set of all channel paths at the ISCC BS is denoted by ${\mathcal{L}_n} = \left\{ {1,2, \cdots ,{L_n}} \right\}$. We adopt a channel model based on the field response, then the channel vector between vehicle $n$ and the ISCC BS is 
\begin{equation}\label{ChannelVector_hn} 
{{\bf{h}}_{n,c}}\left( {t,{\bf{d}}} \right) = {\bf{F}}_{n,c}^H\left( {t,{\bf{d}}} \right){{\bf{G}}_n},
\end{equation}
where ${{\bf{F}}_{n,c}}\left( {t,{\bf{d}}} \right) = \left[ {{{\bf{f}}_{n,c}}\left( {t,{{\bf{d}}_1}} \right), \cdots ,{{\bf{f}}_{n,c}}\left( {t,{{\bf{d}}_M}} \right)} \right]$ $\in {\mathbb{C}^{{L_n} \times M}}$ represents the field-response matrix at the ISCC BS with ${{\bf{f}}_{n,c}}\left( {t,{{\bf{d}}_m}} \right) = {\left[ {{e^{j\frac{{2\pi }}{\lambda }{\rho _{n,1}}\left( {t,{{\bf{d}}_m}} \right)}}, \cdots ,{e^{j\frac{{2\pi }}{\lambda }{\rho _{n,{L_n}}}\left( {t,{{\bf{d}}_m}} \right)}}} \right]^T}$ denoting the field-response vector of the received channel paths between vehicle $n$ and the $m$-th FA at the ISCC BS. In ${{\bf{f}}_{n,c}}\left( {t,{{\bf{d}}_m}} \right)$, ${\rho _{n,l}}\left( {t,{{\bf{d}}_m}} \right)$ is given as follow 
\begin{equation}\label{eqn:rho_nl} 
\begin{array}{*{20}{l}}
{{\rho _{n,l}}\left( {t,{{\mathbf{d}}_m}} \right) = {x_m}\left( t \right)\sin {\theta _{n,l}}\cos {\phi _{n,l}} + {y_m}\left( t \right)\cos {\theta _{n,l}} + } \\ 
{\;\;\;\;\;\;\;\;\;\;\;\;\;\;\;\;\;\;\;\;\;\;{v_n}t\cos \left( {{\theta _{n,l}} - {\theta _{n,v}}} \right).} 
\end{array}
\end{equation}
Equation (\ref{eqn:rho_nl}) denotes the phase difference in the signal propagation for the $l$-th path from vehicle $n$ between the position of the $m$-th FA and the reference point denoted by ${{\mathbf{d}}_0} = {\left[ {0,0} \right]^T}$, where ${\theta _{n,l}}$ and ${\phi _{n,l}}$ represent the elevation and azimuth AoAs for the $l$-th receive path between vehicle $n$ and the ISCC BS. ${v_n}$ is the velocity of vehicle $n$, and ${{\theta _{n,v}}}$ is the movement direction angle of vehicle $n$. The path-response vector is denoted as ${{\mathbf{G}}_n} = {\left[ {{g_{n,1}}, \cdots ,{g_{n,{L_n}}}} \right]^T}$, representing the coefficients of multi-path responses from vehicle $n$ to the reference point in the receive region. The uplink transmission rate from vehicle $n$ to the MEC server at time $t$ is expressed as
\begin{equation}\label{equ:TransmissionRate_Rn} \small
\begin{gathered}
R_n^c\left( t \right) =  \hfill \\
{\log _2}\left( {1 + \frac{{{{\left\| {{\mathbf{w}}_{n,c}^H\left( t \right){{\mathbf{h}}_{n,c}}\left( {t,{\mathbf{d}}} \right)} \right\|}^2}{p_n}}}{{\sum\limits_{k \in \mathcal{N},k \ne n} {{{\left\| {{\mathbf{w}}_{k,c}^H\left( t \right){{\mathbf{h}}_{k,c}}\left( {t,{\mathbf{d}}} \right)} \right\|}^2}{p_k}}  + \left\| {{{\mathbf{w}}_{n,c}}\left( t \right)} \right\|_2^2\sigma _c^2}}} \right). \hfill \\ 
\end{gathered} 
\end{equation}

\subsection{Sensing Model}\label{subsec:SensingModel}
In the proposed FA-assisted vehicle ISCC systems, we utilize the same antenna array at the ISCC BS to simultaneously transmit and receive communication and radar signals. The ISCC BS analyzes the reflected components of communication signals from vehicles to perceive their status, such as position, speed, and road conditions. This approach effectively leverages the reflection of communication signals for radar sensing, achieving dual functionality of the signal. Consequently, the received sensing signal ${{\bf{y}}_s}\left( t \right) \in {\mathbb{C}^{N \times 1}}$ at the ISCC BS can be written as
\begin{equation}\label{equ:SensingReceiveSignal}
{{\bf{y}}_s}\left( t \right) = {\bf{W}}_s^H\left( t \right){{\bf{H}}_s}\left( {t,{\bf{d}}} \right){{\bf{P}}^{{1 \mathord{\left/
				{\vphantom {1 2}} \right.
				\kern-\nulldelimiterspace} 2}}}{\bf{s}}\left( t \right) + {\bf{W}}_s^H\left( t \right){{\bf{n}}_s}\left( t \right),
\end{equation}
where ${{\bf{W}}_s}\left( t \right) = \left[ {{{\bf{w}}_{1,s}}\left( t \right), \cdots ,{{\bf{w}}_{N,s}}\left( t \right)} \right] \in {\mathbb{C}^{M \times N}}$ denotes the receiving steering matrix used in sensing tasks at time $t$ with ${{\bf{w}}_{n,s}}\left( t \right)$ representing the receiving steering vector for the $n$-th vehicle's signal. ${{\bf{n}}_s}\left( t \right) = {\left[ {{n_{1,s}}\left( t \right), \cdots ,{n_{M,s}}\left( t \right)} \right]^T} \sim {\cal C}{\cal N}\left( {{\bf{0}},\sigma _s^2{{\bf{I}}_M}} \right)$ is the AWGN at the ISCC BS at time $t$, modeled as a complex Gaussian vector with zero mean and covariance matrix $\sigma_s^2{{\bf{I}}_M}$, where $\sigma_s^2$ is the sensing noise power. ${{\bf{H}}_s}\left( {t,{\bf{d}}} \right) = \left[ {{{\bf{h}}_{1,s}}\left( {t,{\bf{d}}} \right), \cdots ,{{\bf{h}}_{N,s}}\left( {t,{\bf{d}}} \right)} \right] \in {{\mathbb{C}}^{M \times N}}$ is the target response matrix that describes the reflection paths from all vehicles to the ISCC BS, where ${{\bf{h}}_{n,s}}\left( {t,{\bf{d}}} \right) = {\bf{F}}_{n,s}^H\left( {t,{\bf{d}}} \right){{\bf{G}}_n}$ denotes the sensing channel vector from vehicle $n$ back to the ISCC BS at time $t$, including path loss and phase changes caused by vehicle motion. The matrix, ${{\bf{F}}_{n,s}}\left( {t,{\bf{d}}} \right) = \left[ {{{\bf{f}}_{n,s}}\left( {t,{{\bf{d}}_1}} \right), \cdots ,{{\bf{f}}_{n,s}}\left( {t,{{\bf{d}}_M}} \right)} \right] \in {{\mathbb{C}}^{{L_n} \times M}}$, also referred to as the field-response matrix, captures the impact of the physical environment on the signal from vehicle $n$ as it reflects to the ISCC BS, with each column corresponding to a different antenna element at the ISCC BS. The sensing channel characteristics from vehicle $n$ to the $m$-th antenna element at the ISCC BS at time $t$, including path losses and phase shifts, is described as follows
\begin{equation}\label{equ:fnstdm}
\begin{array}{l}
{{\bf{f}}_{n,s}}\left( {t,{{\bf{d}}_m}} \right) = \\
{\left[ {{\gamma _{n,1}}{e^{j\frac{{2\pi }}{\lambda }{\rho _{n,1}}\left( {t,{{\bf{d}}_m}} \right)}}, \cdots ,{\gamma _{n,{L_n}}}{e^{j\frac{{2\pi }}{\lambda }{\rho _{n,{L_n}}}\left( {t,{{\bf{d}}_m}} \right)}}} \right]^T},
\end{array}
\end{equation}
where ${\gamma _{n,l}} = {\alpha _{n,l}}{e^{j\frac{{2\pi }}{\lambda }{\tau _{n,l}}}}$ is the combined effect of reflection coefficient ${\alpha _{n,l}}$ and time delay ${\tau _{n,l}}$ of vehicle $n$ at the $l$-th path. The sensing rate from vehicle $n$ to the MEC server at time $t$ is given by
\begin{equation}\label{equ:SensingRate_Rns} 
\begin{array}{l}
R_n^s\left( t \right) = \\
{\log _2}\left( {1 + \frac{{{{\left\| {{\bf{w}}_{n,s}^H\left( t \right){{\bf{h}}_{n,s}}\left( {t,{\bf{d}}} \right)} \right\|}^2}{p_n}}}{{\sum\limits_{k \in {\cal N},k \ne n} {{{\left\| {{\bf{w}}_{k,s}^H\left( t \right){{\bf{h}}_{k,s}}\left( {t,{\bf{d}}} \right)} \right\|}^2}{p_k}}  + \left\| {{{\bf{w}}_{n,s}}\left( t \right)} \right\|_2^2\sigma _s^2}}} \right).
\end{array}
\end{equation}

\section{Integrated Latency Optimization Problem Formulation} \label{sec:LatencyOptimizationProblemFormulation}
Vehicles often face limited computing resources, making it challenging to perform complex tasks locally. To address this issue, the use of the MEC server at the ISCC BS has been proposed to enhance the computational capabilities of vehicles. All vehicles run the federated learning (FL) training tasks, tasks are offloaded to the MEC server at the ISCC BS due to limited computing resources of vehicles. We employ the stochastic gradient descent (SGD) optimization algorithm to train the FL model on the MEC server. 

We consider that vehicle $n$ offloads all datasets to the MEC server, where the communication model is subsequently trained. The communication model tasks may include channel estimation, beamforming optimization, power control, and interference management. Consequently, the transmission latency involved in offloading datasets from vehicle $n$ to the MEC server is determined as
\begin{equation}\label{equ:Tnc_off}
T_{n,c}^{off,\left( t \right)} = \frac{{D_{n,c}^{\left( t \right)}}}{{R_n^c\left( t \right)}},
\end{equation}
where $D_{n,c}^{\left( t \right)}$ denotes the size of the datasets offloaded from vehicle $n$ to the MEC server during time slot $t$. Then, the execution time of communication model training for vehicle $n$ on the MEC server is expressed as
\begin{equation}\label{equ:Tnc_exe}
T_{n,c}^{exe,\left( t \right)} = \frac{{{C_M}D_{n,c}^{\left( t \right)}\varpi _M^{\left( t \right)}\iota _M^{\left( t \right)}}}{{f_{n,c}^{\left( t \right)}}},
\end{equation}
where ${C_M}$ is regarded as the number of CPU cycles required to process a single data sample for the MEC server, ${\iota _M^{\left( t \right)}}$ represents the number of iterations of the SGD algorithm on the MEC server, and $\varpi _M^{\left( t \right)} \in \left( {0,1} \right]$ indicates the mini-batch size ratio on the MEC server. ${f_{n,c}^{\left( t \right)}}$ refers to the CPU frequency of vehicle $n$ from the MEC server at time $t$ for performing communication tasks. The vector ${\mathbf{f}}_c^{\left( t \right)} = \left[ {f_{1,c}^{\left( t \right)},f_{2,c}^{\left( t \right)}, \cdots ,f_{N,c}^{\left( t \right)}} \right] \in {\mathbb{C}^{1 \times N}}$ represents the CPU frequencies for all  vehicles in the network during time slot $t$ dedicated to communication tasks. Then, the total communication latency of vehicle $n$ is calculated as
\begin{equation}\label{equ:Tnc_total}
T_{n,c}^{\left( t \right)} = T_{n,c}^{off,\left( t \right)} + T_{n,c}^{exe,\left( t \right)}.
\end{equation}

In addition to communication, the proposed FA-enabled vehicle ISCC system must also handle extensive sensing tasks to support functions like environment perception and obstacle detection. In vehicular networks, sensing tasks are crucial for ensuring safety and enabling advanced functionalities. These tasks often require substantial computational resources that exceed the capabilities of individual vehicle. Similar to communication tasks, vehicles can offload their sensing datasets to the MEC server at the ISCC BS, where more powerful computational resources are available.

We consider that vehicle $n$ offloads all sensing datasets to the MEC server, where the sensing model is subsequently processed. The sensing model tasks may include object detection, lane detection, traffic sign recognition, and environmental mapping. As a result, the transmission latency required to transfer the sensing datasets from vehicle $n$ to the MEC server is expressed as
\begin{equation}\label{equ:Tns_off}
T_{n,s}^{off,\left( t \right)} = \frac{{D_{n,s}^{\left( t \right)}}}{{R_n^s\left( t \right)}},
\end{equation}
where $D_{n,s}^{\left( t \right)}$ denotes the size of the sensing datasets offloaded from vehicle $n$ to the MEC server during time slot $t$. Then, the execution time for processing sensing tasks for vehicle $n$ on the MEC server is expressed as

\begin{equation}\label{equ:Tns_exe}
T_{n,s}^{exe,\left( t \right)} = \frac{{{C_M}D_{n,s}^{\left( t \right)}\varpi _M^{\left( t \right)}\iota _M^{\left( t \right)}}}{{f_{n,s}^{\left( t \right)}}},
\end{equation}
where ${f_{n,s}^{\left( t \right)}}$ refers to the CPU frequency of vehicle $n$ from the MEC server at time $t$ for executing sensing tasks. The vector ${\mathbf{f}}_s^{\left( t \right)} = \left[ {f_{1,s}^{\left( t \right)},f_{2,s}^{\left( t \right)}, \cdots ,f_{N,s}^{\left( t \right)}} \right] \in {\mathbb{C}^{1 \times N}}$ captures the CPU frequencies assigned for sensing tasks across all vehicles in the network during time slot $t$. Then, the total sensing latency of vehicle $n$ is calculated as
\begin{equation}\label{equ:Tns_total}
T_{n,s}^{\left( t \right)} = T_{n,s}^{off,\left( t \right)} + T_{n,s}^{exe,\left( t \right)}.
\end{equation}

Based on the derived latencies for communication and sensing, we can compute the overall latency for each vehicle and the entire system. For vehicle $n$ in time slot $t$, the total latency is given by
\begin{equation}\label{equ:Tntoal}
T_n^{total,\left( t \right)} = T_{n,c}^{\left( t \right)} + T_{n,s}^{\left( t \right)}.
\end{equation}
Summing up the total latency for all vehicles in the network provides the system-wide total latency as follows
\begin{equation}\label{equ:Ttoal}
{T^{total,\left( t \right)}} = \sum\nolimits_{n \in \mathcal{N}} {T_n^{total,\left( t \right)}}.
\end{equation}

To optimize the performance of the FA-enabled vehicle ISCC system, it is essential to minimize the total latency experienced by all vehicles in the network. This involves determining the optimal resource allocation for communication and sensing tasks, represented by the variables ${\mathbf{d}}\left( t \right)$ for antenna positions, ${{\mathbf{W}}_c}\left( t \right)$ and ${{\mathbf{W}}_s}\left( t \right)$ for communication and sensing receive combining matrices, ${\mathbf{f}}_c^{\left( t \right)}$ and ${\mathbf{f}}_s^{\left( t \right)}$ for the CPU frequencies allocated to communication and sensing tasks, respectively. Therefore, we formulate the following optimization problem, denoted as $\mathcal{P}1$, to minimize the total latency ${T^{total,\left( t \right)}}$ across the network as follows

\begin{equation}\label{eqn:problem_P1}
\begin{array}{l}
\mathcal{P}1:\mathop {\min }\limits_{{\mathbf{d}}\left( t \right),{{\mathbf{W}}_c}\left( t \right),{{\mathbf{W}}_s}\left( t \right),{\mathbf{f}}_c^{\left( t \right)},{\mathbf{f}}_s^{\left( t \right)}} {T^{total,\left( t \right)}}\\
\;\;\;\;\;\;\;\;{\rm{s}}.{\rm{t}}.\;{\rm{C1}}:{{\bf{d}}_m}\left( t \right) \in {\mathcal{D}_r},\forall m \in \mathcal{M},\\
\;\;\;\;\;\;\;\;\;\;\;\;\;\;{\rm{C2}}:\left\| {{{\bf{d}}_m}\left( t \right) - {{\bf{d}}_\kappa\left( t \right) }} \right\| \ge {d_0},m \ne \kappa ,\forall m \in \mathcal{M},\\
\;\;\;\;\;\;\;\;\;\;\;\;\;\;{\rm{C3}}:T_{n,c}^{\left( t \right)} \le T_{th}^c,\forall n \in \mathcal{N},\\
\;\;\;\;\;\;\;\;\;\;\;\;\;\;{\rm{C4}}:T_{n,s}^{\left( t \right)} \le T_{th}^s,\forall n \in \mathcal{N},\\
\;\;\;\;\;\;\;\;\;\;\;\;\;\;{\rm{C5}}:\sum\nolimits_{n \in {\cal N}} {\left( {f_{n,c}^{\left( t \right)} + f_{n,s}^{\left( t \right)}} \right) \le {f_{th}}}, 
\end{array}
\end{equation}
where constraint ${\rm{C1}}$ means the position of FA at time $t$ within the predefined region ${\mathcal{D}_r}$. Constraint ${\rm{C2}}$ ensures that the distance between any two antennas at time $t$ is at least ${d_0}$. Constraint ${\rm{C3}}$ states that the communication latency $T_{n,c}^{\left( t \right)}$ for vehicle $n$ at time $t$ must not exceed the threshold $T_{th}^c$. Similarly, constraint ${\rm{C4}}$ specifies that the sensing latency $T_{n,s}^{\left( t \right)}$ for vehicle $n$ at time $t$ must not exceed the threshold $T_{th}^s$. Lastly, constraint ${\rm{C5}}$ states that the sum of the CPU frequencies allocated to communication and sensing for all vehicles at time $t$ must not exceed the threshold ${f_{th}}$. By analyzing the problem $\mathcal{P}1$, $\mathcal{P}1$ is a mixed discrete and non-convex optimization problem, which is also well known as an NP-hard problem.

\section{Communication, Sensing, and Computing Resource Allocation} \label{sec:ISCCCResourceAllocation}
To solve the final solutions, we decompose the original problem $\mathcal{P}1$ into three subproblems. We first analyze the sub-problem $\mathcal{P}1.1$, which aims to minimize the total system latency ${T^{total,\left( t \right)}}$ by optimizing the allocation of computing resources $\left( {{\mathbf{f}}_c^{\left( t \right)},{\mathbf{f}}_s^{\left( t \right)}} \right)$ during communication and sensing. In time slot $t$, the sub-problem $\mathcal{P}1.1$ can be specifically expressed as follows
\begin{equation}\label{eqn:problem_P1.1}
\begin{gathered}
\mathcal{P}1.1:\mathop {\min }\limits_{{\mathbf{f}}_c^{\left( t \right)},{\mathbf{f}}_s^{\left( t \right)}} {T^{total,\left( t \right)}} \;\;{\text{s}}.{\text{t}}.\;{\text{C3}} - {\text{C5}}, 0 \le t \le {\cal T}. \hfill \\  
\end{gathered} 
\end{equation}
By analyzing this subproblem, the subproblem ${\cal P}1.1$ is a convex optimization problem. Traditional optimization methods such as interior point method, gradient projection method, and Lagrange dual method can be used to find the optimal CPU frequency of ${\cal P}1.1$.

Given the computing resource strategies $\left( {{\mathbf{f}}_c^{\left( t \right)},{\mathbf{f}}_s^{\left( t \right)}} \right)$ and the antenna positioning strategies ${\mathbf{d}}\left( t \right)$, the objective function ${T^{total,\left( t \right)}}$ is determined by the receive combining matrix $\left( {{{\mathbf{W}}_c}\left( t \right),{{\mathbf{W}}_s}\left( t \right)} \right)$. By optimizing the receive combining matrix $\left( {{{\mathbf{W}}_c}\left( t \right),{{\mathbf{W}}_s}\left( t \right)} \right)$ during communication and sensing, the sub-problem $\mathcal{P}1.2$ can be specifically formulated as follows
\begin{equation}\label{eqn:problem_P1.2}
\begin{gathered}
\mathcal{P}1.2:\mathop {\min }\limits_{{{\mathbf{W}}_c}\left( t \right),{{\mathbf{W}}_s}\left( t \right)} {T^{total,\left( t \right)}} \;\;\;{\text{s}}.{\text{t}}.\;{\text{C3}},{\text{C4}},0 \le t \le {\cal T}. \hfill \\ 
\end{gathered} 
\end{equation}
To simplify the structure of problem $\mathcal{P}1.2$ and speed up the solution process, we decouple the problem $\mathcal{P}1.2$ into the following two sub-problems
\begin{equation}\label{eqn:problem_P1.2a}
\begin{gathered}
\mathcal{P}1.2\left( a \right):\mathop {\min }\limits_{{{\mathbf{W}}_c}\left( t \right)} {T^{total,\left( t \right)}} \;\;\; {\text{s}}.{\text{t}}.\;{\text{C3,}}\hfill \\
\end{gathered} 
\end{equation}
\begin{equation}\label{eqn:problem_P1.2b}
\begin{gathered}
\mathcal{P}1.2\left( b \right):\mathop {\min }\limits_{{{\mathbf{W}}_s}\left( t \right)} {T^{total,\left( t \right)}} \;\;\;{\text{s}}.{\text{t}}.\;{\text{C4.}} \hfill \\
\end{gathered}
\end{equation}

Let us first focus on sub-problem $\mathcal{P}1.2\left( a \right)$. By observing this subproblem, the problem $\mathcal{P}1.2\left( a \right)$ is a non-convex optimization problem. By introducing matrix variables, we reformulate the original problem $\mathcal{P}1.2\left( a \right)$ as a semidefinite programming (SDP) problem. To represent the received combining vector as a matrix, we introduce the following variables

\begin{equation}\label{eqn:MatrixVariables}
\left\{ \begin{gathered}
{{{\mathbf{\tilde W}}}_{n,c}}\left( t \right) = {{\mathbf{w}}_{n,c}}\left( t \right){\mathbf{w}}_{n,c}^H\left( t \right) \in {\mathbb{C}^{M \times M}}, \hfill \\
{{{\mathbf{\tilde H}}}_{n,c}}\left( t \right) = {{\mathbf{h}}_{n,c}}\left( t \right){\mathbf{h}}_{n,c}^H\left( t \right) \in {\mathbb{C}^{M \times M}}. \hfill \\ 
\end{gathered}  \right.
\end{equation}
Then, we have ${{{\mathbf{\tilde W}}}_c}\left( t \right) = \left[ {{{{\mathbf{\tilde W}}}_{1,c}}\left( t \right), \cdots, {{{\mathbf{\tilde W}}}_{N,c}}\left( t \right)} \right]$ and ${{{\mathbf{\tilde H}}}_c}\left( t \right) = \left[ {{{{\mathbf{\tilde H}}}_{1,c}}\left( t \right), \cdots, {{{\mathbf{\tilde H}}}_{N,c}}\left( t \right)} \right]$. We replace the received combining vector ${{\mathbf{w}}_{n,c}}\left( t \right)$ and the channel vector ${{\mathbf{h}}_{n,c}}\left( t \right)$ in $T_{n,c}^{off,\left( t \right)}$ with the matrix variables ${{{\mathbf{\tilde W}}}_{n,c}}\left( t \right)$ and ${{{\mathbf{\tilde H}}}_{n,c}}\left( t \right)$. 
Using the trace and eigenvalue properties of the matrix, we can get a lower bound on the transmission latency $T_{n,c}^{off,\left( t \right)}$ involved in offloading datasets of communication tasks from vehicle $n$ to the MEC server
\begin{equation}\label{eqn:Tncoff_v3}
\begin{gathered}
\tilde T_{n,c}^{off,\left( t \right)} =  \hfill \\
\frac{{D_{n,c}^{\left( t \right)}}}{{{{\log }_2}\left( {1 + \frac{{{\lambda _{\max }}\left( {{{{\mathbf{\tilde H}}}_{n,c}}\left( t \right)} \right){\text{Tr}}\left( {{{{\mathbf{\tilde W}}}_{n,c}}\left( t \right)} \right){p_n}}}{{\sum\limits_{k \in \mathcal{N},k \ne n} {{\lambda _{\min }}\left( {{{{\mathbf{\tilde H}}}_{k,c}}\left( t \right)} \right){\text{Tr}}\left( {{{{\mathbf{\tilde W}}}_{k,c}}\left( t \right)} \right){p_k}}  + \operatorname{Tr} ({{{\mathbf{\tilde W}}}_{n,c}}\left( t \right))\sigma _c^2}}} \right)}}, \hfill \\ 
\end{gathered}
\end{equation}
where ${{\lambda _{\max }}\left( {{{{\mathbf{\tilde H}}}_{n,c}}\left( t \right)} \right)}$ and ${{\lambda _{\min }}\left( {{{{\mathbf{\tilde H}}}_{n,c}}\left( t \right)} \right)}$ are the largest and smallest eigenvalues of matrix ${{{{\mathbf{\tilde H}}}_{n,c}}\left( t \right)}$, respectively. By introducing the matrix variables and inequality transformation, the original sub-problem $\mathcal{P}1.2\left( a \right)$ can be equivalent to 

\begin{equation}\label{eqn:problem_P1.2a_v2}
\begin{gathered}
\mathcal{P}1.2{\left( a \right)^\prime }: \hfill \\
\mathop {\min }\limits_{{{{\mathbf{\tilde W}}}_c}\left( t \right)} \sum\nolimits_{n \in \mathcal{N}} {\left( {\tilde T_{n,c}^{off,\left( t \right)} + T_{n,c}^{exe,\left( t \right)} + T_{n,s}^{\left( t \right)}} \right)}  \hfill \\
{\text{s}}.{\text{t}}.\;{\text{C3'}}:\tilde T_{n,c}^{off,\left( t \right)} + T_{n,c}^{exe,\left( t \right)} - T_{th}^c \le 0,\forall n \in \mathcal{N}, \hfill \\
\;\;\;\;\;\,{\text{C6}}:{{{\mathbf{\tilde W}}}_{n,c}}\left( t \right) \underset{\raise0.3em\hbox{$\smash{\scriptscriptstyle}$}}{ \succeq } 0 ,\forall n \in \mathcal{N}, \hfill \\ 
\end{gathered}
\end{equation}
where constraint ${\text{C6}}$ ensures the semi-positive definiteness of the matrix ${{{\mathbf{\tilde W}}}_{n,c}}$. By observing this subproblem, the problem  $\mathcal{P}1.2{\left( a \right)^\prime }$ is a convex optimization problem. Then, we solve the equivalent problem $\mathcal{P}1.2{\left( a \right)^\prime }$ using a standard SDP solver such as the Matlab toolbox CVX. Similarly, the sub-problem $\mathcal{P}1.2\left( b \right)$ can also be approximated into a convex problem. 

\begin{algorithm}[!t] \scriptsize
	\caption{IDPSO-based Alternating Iterative Algorithm for Solving Problem $\mathcal{P}1$ \scriptsize} \label{Algorithm_Overall}
	\setcounter{algorithm}{1}
	\begin{algorithmic}[1]
		\STATE {\textbf{Input:} The initial data set $\left( M, N, L,\sigma _c^2,\sigma _s^2 \right., {\theta _{n,l}},{\phi _{n,l}}, {\theta _{n,v}},{v_n}$, ${g_{n,l}},{\alpha _{n,l}},{\tau _{n,l}}, D_{n,c}^{\left( 0 \right)},D_{n,s}^{\left( 0 \right)}$ for $n \in {\cal N}$, $m \in {\cal M}$, and $l \in {{\cal L}_n}$.}
		\STATE {\textbf{Initialization:} Initialize target variables ${\bf{d}}\left( 0 \right)$, ${{\mathbf{W}}_c}\left( 0 \right)$, ${{\mathbf{W}}_s}\left( 0 \right)$, ${\mathbf{f}}_c^{\left( 0 \right)}$, and ${\mathbf{f}}_s^{\left( 0 \right)}$.}
		\FOR {Time slot $t=0$ to $\cal T$}
		\FOR {Iteration $i=1$ to $I$}
		\STATE {Fix $\left( {{{\bf{W}}_c}\left( t \right),{{\bf{W}}_s}\left( t \right),{\bf{d}}\left( t \right)} \right)$, compute the computing resource allocation strategies $\left( {{\mathbf{f}}_c^{\left( t \right)},{\mathbf{f}}_s^{\left( t \right)}} \right)$ by solving sub-problem $\mathcal{P}1.1$ using interior point algorithm.}
		\STATE {Fix $\left( {{\bf{f}}_c^{\left( t \right)},{\bf{f}}_s^{\left( t \right)},{\bf{d}}\left( t \right)} \right)$, compute the receive combining strategies $\left( {{{\mathbf{W}}_c}\left( t \right),{{\mathbf{W}}_s}\left( t \right)} \right)$ by solving sub-problem $\mathcal{P}1.2$ using SDP-based alternating iterative algorithm.}
		\STATE {Fix $\left( {{\bf{f}}_c^{\left( t \right)},{\bf{f}}_s^{\left( t \right)},{{\bf{W}}_c}\left( t \right),{{\bf{W}}_s}\left( t \right)} \right)$, compute the antenna positioning strategies ${\mathbf{d}}\left( t \right)$ by solving sub-problem $\mathcal{P}1.3$ using PSO algorithm.}
		\ENDFOR
		\ENDFOR
		\STATE {\textbf{Output:} The optimal solutions for $\left( {\mathbf{d}}, {{\mathbf{W}}_c}, {{\mathbf{W}}_s}, {\mathbf{f}}_c, {\mathbf{f}}_s \right)$.}
	\end{algorithmic}
\end{algorithm}

Given the computing resource strategies $\left( {{\mathbf{f}}_c^{\left( t \right)},{\mathbf{f}}_s^{\left( t \right)}} \right)$ and the receive combining strategies $\left( {{{\mathbf{W}}_c}\left( t \right),{{\mathbf{W}}_s}\left( t \right)} \right)$, the objective function ${T^{total,\left( t \right)}}$ is determined by the antenna positioning matrix ${{\mathbf{d}}\left( t \right)}$. By optimizing the antenna positioning matrix $\left( {{{\mathbf{W}}_c}\left( t \right),{{\mathbf{W}}_s}\left( t \right)} \right)$ during communication and sensing, the sub-problem $\mathcal{P}1.3$ can be specifically given by
\begin{equation}\label{eqn:problem_P1.3}
\begin{gathered}
\mathcal{P}1.3:\mathop {\min }\limits_{{\mathbf{d}}\left( t \right)} {T^{total,\left( t \right)}} \hfill \\
\;\;\;\;\;\;\;\;\;\;{\text{s}}.{\text{t}}.\;{\text{C1}} - {\text{C4}},0 \le t \le {\cal T}. \hfill \\ 
\end{gathered} 
\end{equation}
The sub-problem $\mathcal{P}1.3$ is a non-convex optimization problem. We employ the particle swarm optimization (PSO) algorithm \cite{zuo2024fluid} as an effective method to solve the problem $\mathcal{P}1.3$.

To solve problem $\mathcal{P}1$, we decompose it into three sub-problems: computing resource optimization problem $\mathcal{P}1.1$, receive combining optimization problem $\mathcal{P}1.2$, and antenna positioning optimization problem $\mathcal{P}1.3$. Each sub-problem is tackled using different optimization techniques. Based on the above analysis, we design a mixed interior point, SDP, and PSO (IDPSO) alternating iterative algorithm to solve the overall problem $\mathcal{P}1$. The proposed IDPSO-based alternating iterative algorithm integrates three specialized algorithms to address the sub-problems effectively. The computational resource allocation is handled using an interior point method, the receive combining optimization leverages the SDP-based alternating iterative approach, and the antenna positioning optimization employs the PSO algorithm. By iteratively solving each sub-problem and updating the variables accordingly, the IDPSO-based alternating iterative algorithm converges towards the optimal solutions for the entire problem $\mathcal{P}1$. For the detailed process, the proposed IDPSO-based alternating iterative algorithm is summarized in Algorithm~\ref{Algorithm_Overall}. The computational
complexity of Algorithm~\ref{Algorithm_Overall} is $\mathcal{O}\left( {\mathcal{T}I{N^{6.5}} + {M^3}} \right)$, where $\mathcal{T}$ indicates the number of iterations of the outer loop, and $I$ is the number of iterations of the inner loop.
\begin{table}[!t] \scriptsize 
	\centering
	\caption {Simulation Parameters} 
	\begin{tabular}{l|l}
		\hline \hline
		\textbf{Parameter} & \textbf{Value}          
		\\ \hline
		Number of FAs at the ISCC BS, $M$                           
		&
		$4$
		\\ \hline
		Number of vehicles, $N$                                                       
		&
		$3$  
		\\ \hline
		Number of channel paths for each vehicle, $L_n$ 
		&  
		$3$ 
		\\ \hline
		Carrier wavelength, $\lambda$
		&
		$0.1$ m
		\\ \hline
		Length of sides of receive area, $A$
		&
		$1.5\lambda$
		\\ \hline
		Minimum inter-FA distance, $d_0$
		&
		$\lambda$
		\\ \hline
		Data size of vehicle $n$, ${D_{n,c}^{\left( t \right)}} = {D_{n,s}^{\left( t \right)}}$
		& 
		$\left[ {0.5 - 2} \right]$ KB
		\\ \hline
		Elevation and azimuth AoAs, ${\theta _{n,l}}={\phi _{n,l}}$
		&
		$\left[ { - \frac{\pi }{2},\frac{\pi }{2}} \right]$
		\\ \hline
		Maximum CPU frequency of MEC server, $f_{th}$                                                           
		& 
		$100$ GHz 
		\\ \hline
		Transmit power for each vehicle, ${p_n}$                                                
		& 
		$30$ dBm 
		\\ \hline
		Noise power spectrum density, ${{\sigma_{c} ^2}} = {{\sigma_{s} ^2}}$                                                
		& 
		$-174$ dBm/Hz
		\\ \hline
		Reflection coefficient, ${\alpha _{n,l}}$
		&
		0.5
		\\ \hline
	\end{tabular}
	\label{SimulationParameters}
\end{table}
\begin{figure}[!t]
	\centering    
	\includegraphics[width=0.75\linewidth]{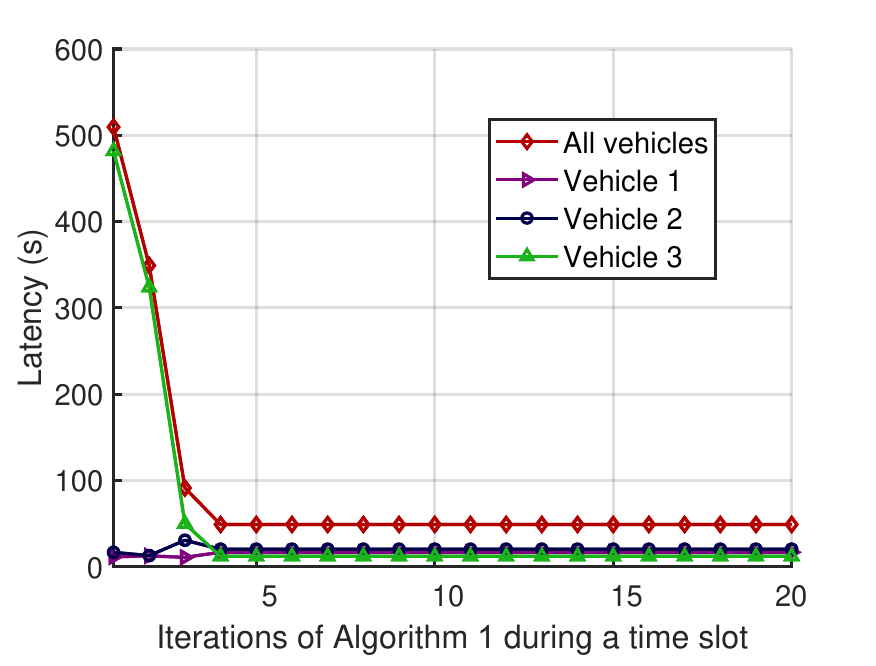} 
	\caption{Convergence performance of the proposed Algorithm~\ref{Algorithm_Overall}.}
	\label{fig:Convergence}
\end{figure}

\begin{figure}[!t]
	\centering
	\includegraphics[width=0.75\linewidth]{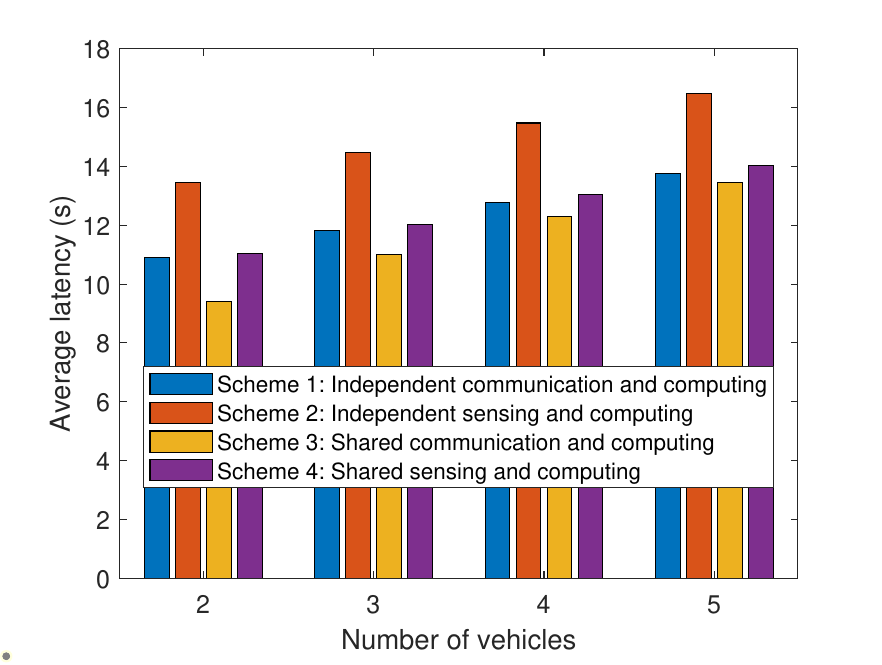}
	\caption{Average latency in different schemes.}
	\label{fig:AverageLatency}
\end{figure}

\section{Numerical Results}\label{sec:NumericalResults}
For the sake of illustration, let us consider a simple example. In the FA-enabled vehicle ISCC network, there are 3 vehicles, the ISCC BS is equipped with 4 antennas, and the number of channel paths for each vehicle is 3. The main simulation parameters are listed in TABLE~\ref{SimulationParameters}. Fig.~\ref{fig:Convergence} demonstrates the convergence performance of the proposed IDPSO-based alternating iterative algorithm during a time slot. As shown in Fig.~\ref{fig:Convergence}, after more than 5 iterations, the latencies of all vehicles and the total latency reach a stable state. Therefore, we can conclude that when $N=3$, $M=4$, and $L_n=3$, Algorithm~\ref{Algorithm_Overall} has the fast convergence performance.

\begin{figure}[!t]
	\centering
	\includegraphics[width=0.75\linewidth]{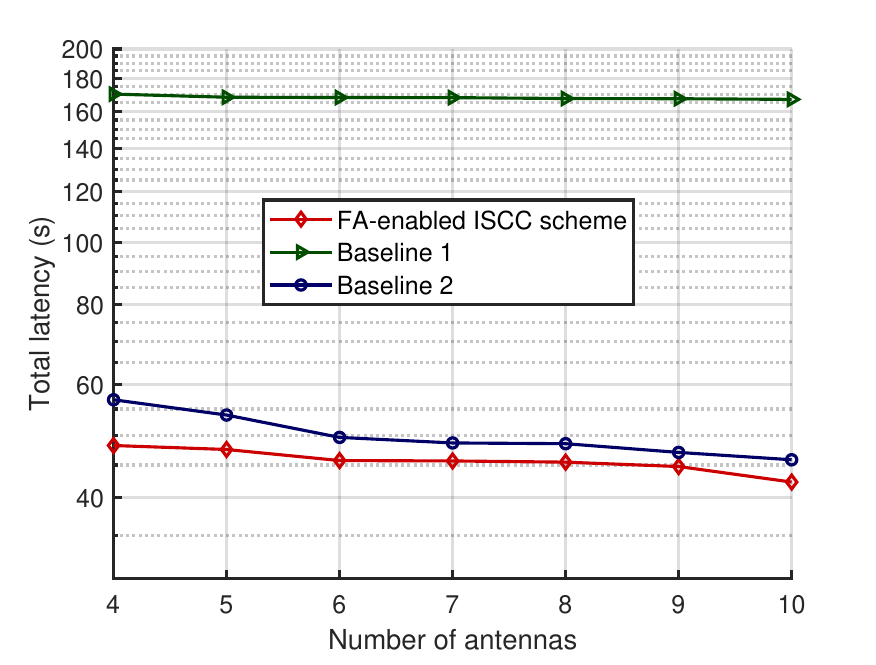}
	\caption{Comparisons of FA-Enabled ISCC and baseline schemes.}
	\label{fig:TotalLatency_MN}
\end{figure}

To demonstrate the advantages of our proposed FA-enabled vehicle ISCC scheme, we compare it with two other schemes in which communication and sensing are relatively independent. Scheme 1 represents independent communication and computing, while scheme 2 focuses on independent sensing and computing. In scheme 1 and scheme 2, communication and sensing employ independent and equal system resources, respectively. Scheme 3 is the communication and computing aspects of our proposed scheme, and scheme 4 focuses on the sensing and computing part of our proposed scheme. In both scheme 3 and scheme 4, communication and sensing share system resources. As shown in Fig.~\ref{fig:AverageLatency}, we can observe that as the number of vehicles increases, the average latency of all schemes increase. But the total latency of scheme 3 and scheme 4 is always less than that of scheme 1 and scheme 2. This is because scheme 1 and scheme 2 adopt system resources independently, which leads to resource waste and increase system delay. However, our proposed scheme 3 and scheme 4 can fully adopt all resources. 

Fig.~\ref{fig:TotalLatency_MN} illustrates the comparison of the proposed FA-enabled vehicle ISCC scheme with two baseline schemes in terms of total latency \cite{zuo2024fluid}. In baseline 1, all vehicles never offload the communication and sensing tasks to the MEC server, and all antennas on the ISCC BS are fixed at the origin. Baseline 2 allows the vehicle to offload all communication and sensing tasks to the MEC server, but the antenna positions on the ISCC BS are also fixed at the origin. In Fig.~\ref{fig:TotalLatency_MN}, we set $N=3$ and $L_n=3$. Observing Fig.~\ref{fig:TotalLatency_MN}, we find that that baseline 1 has the highest total latency, mainly because the vehicle relies only on local computation, resulting in longer latency. In contrast, the FA-enabled vehicle ISCC scheme exhibits a lower total latency than baseline 2. The numerical results also show that the total latency of each scheme decreases with the increment of the number of antennas. Compared with baseline 1 and baseline 2, which have fixed antenna positions, the FA-enabled vehicle ISCC scheme shows significant improvements. By jointly optimizing all FA positions in continuous space, this scheme enhances communication, sensing, and computation performance, resulting in a substantial reduction in total system latency. 
 
\section{Conclusion} \label{sec:Conclusion}
We have designed a FA-enabled ISCC system specifically applied in vehicular networks.  We began by introducing the detailed models of the communication and sensing processes in the FA-enabled vehicle ISCC system. Following this, we formulated an integrated latency optimization problem to jointly optimize the computing resources, the receive combining matrices, and the antenna positions. To tackle this complex problem, we decomposed it into three sub-problems. We also developed a mixed IDPSO-based alternating iterative algorithm to solve the overall problem effectively. Through numerical simulations, we verified the rapid convergence of the proposed IDPSO-based algorithm. The FA-enabled ISCC system demonstrated a significant reduction in total latency compared to baseline schemes, highlighting its capability to effectively utilize system resources.

\section{ACKNOWLEDGEMENT}
This work was supported in part by the National Natural Science Foundation of China (NSFC) under Grants 62301279, 62401137, 62401640, 62102196, 62372248, 62172236, and by the Guangdong Basic and Applied Basic Research Foundation under Grant 2023A1515110732. This work was supported in part by the Postdoctoral Fellowship Program of the China Postdoctoral Science Foundation (CPSF) under Grant BX20230065, and by the National Science Foundation for Excellent Young Scholars of Jiangsu Province under Grant No. BK20220105.
\bibliographystyle{IEEEtran}
\bibliography{IEEEabrv,FAISAC}
\end{document}